\documentclass[11pt]{article}
\usepackage{Blois,epsfig}

\def\be{\begin{equation}}
\def\ee{\end{equation}}
\def\bea{\begin{eqnarray}}
\def\eea{\end{eqnarray}}

\begin{document}
\vspace*{2cm}
\begin{center}
\Large{\textbf{XIth International Conference on\\ Elastic and Diffractive Scattering\\ Ch\^{a}teau de Blois, France, May 15 - 20, 2005}}
\end{center}

\vspace*{2cm}
\title{HADRONIC COSMIC RAY INTERACTIONS NEAR THE LHC ENERGY REGION
 AND IN THE UHE DOMAIN OF GIANT EAS}

\author{ J.N. CAPDEVIELLE }

\address{APC, College de France, 11 Place M.Berthelot,\\
75231 Paris  Cedex 01, France}

\maketitle\abstracts{
The fluctuations of $\gamma$ ray families simulated with CORSIKA in the energy
region
3$\times$10$^{15}$-10$^{17}$eV on the basis of
  standard $Ln s$ collider physics exhibits alignments of secondaries 
in the stratosphere
 and at ground level. 
The remarkable event registrated on the Concorde doesn't
 fit well however those cases ; 
The possible hints of new mechanisms, especially the valence 
diquark breaking, are considered.
 Observing that the extrapolation of the original cosmic ray primary
 spectrum derived from the size spectrum
 measured in the Akeno classical EAS array coincides 
with the spectrum measured recently by
 the Hires Stereo experiment, we point out a possible overestimation 
of the primary energy in inclined showers of the surface arrays like AGASA.}

\section{Remarkable events with coplanar emission} \label{subsec:prod} 
   The attention on coplanar emission  was motivated by the events 
recorded in Pamir X ray chamber~\cite{boris}, in Kambala, in Kascade, 
but also in balloon experiments and in the low stratosphere 
with  high resolution X ray emulsion chambers in the Concorde~\cite{cap88}; 
the geometrical criteria used to select an alignment treat directly 
the coordinates of the individual $\gamma$'s, 
either the linear coefficient~\cite{jn01}:
\begin{equation}
  r = \frac{\sum_{i}^{n}(x_{i} - \bar{x})(y_{i} - \bar{y})}
           {\sqrt{\sum_{i}^{n}(x_{i} - \bar{x})^2}
      \sqrt{\sum_{i}^{n}(y_{i} - \bar{y})^2}}
\label{eq:correlation}
\end{equation}
 or the parameter $\lambda_n$  $[1]$ defined as:
\begin{equation}
   \lambda_n = \frac{\sum_{i \neq j \neq k}^{n}
               \cos 2 \varphi_{ij}^{k}} {n (n-1) (n-2)}
\label{eq:lambda}
\end{equation}

\noindent where $\varphi_{ij}^{k}$ is the angle between the straight
lines joining the $i^{\rm th}$ and $j^{\rm th}$ particles to the
$k^{\rm th}$ one ($0\le \varphi_{ij}^ k \le \pi$).
Alignments have been easily obtained, independantly on the models 
(as far as they have similar
transverse momenta distribution),
and practically on the  primary mass and energy~\cite{atta}.
\section{The Concorde event near $10^{16}$ eV}
 Tracing back  the aligned events produced with CORSIKA 
at Concorde altitude, we have observed that
 an unbalanced $p_{t}$ received in the first collision 
on an energetic $\pi^{0}
$ could produce an alignment when the cascade starts 10km above the chamber.
\begin{table}[ht]
\begin{center}
\begin{tabular}{lccccc}
\hline
   & DPMJET & HDPM & QGSJET & SIBYLL & VENUS \\
\hline
$|r| \ge 0.94$       & 0.7 & 1.5 & 0.6 & 0.5 & 0.9 \\
\hline
$\lambda_4 \ge 0.8$  & 7.4 & 8.0 & 7.4 & 7.4 & 7.1 \\
\hline
\end{tabular}
\end{center}
\caption{Calculated fractions (\%) 
of the aligned events 
with at least 4
$\gamma$-rays ($\gamma$+e$^{\pm}$) above 10 TeV 
for with $|r| \ge 0.94$
5(first row) and $\lambda_4 \ge 0.8$ (second row)
 for different high
energy hadronic interaction models.
\label{tab:concorde} }
\end{table}
Among the 211 $\gamma$'s of the JF2aF2 Concorde event, we show on a lego plot the energy deposited in a plane perpendicular to the axis by 34 $\gamma$'s (about one half of the total energy deposited, i.e. $1600$ TeV).
\begin{figure}[h]
\includegraphics*[width=0.48\textwidth,bbllx=55pt,bblly=144pt,bburx=535pt,bbury=590pt,angle=0,clip]{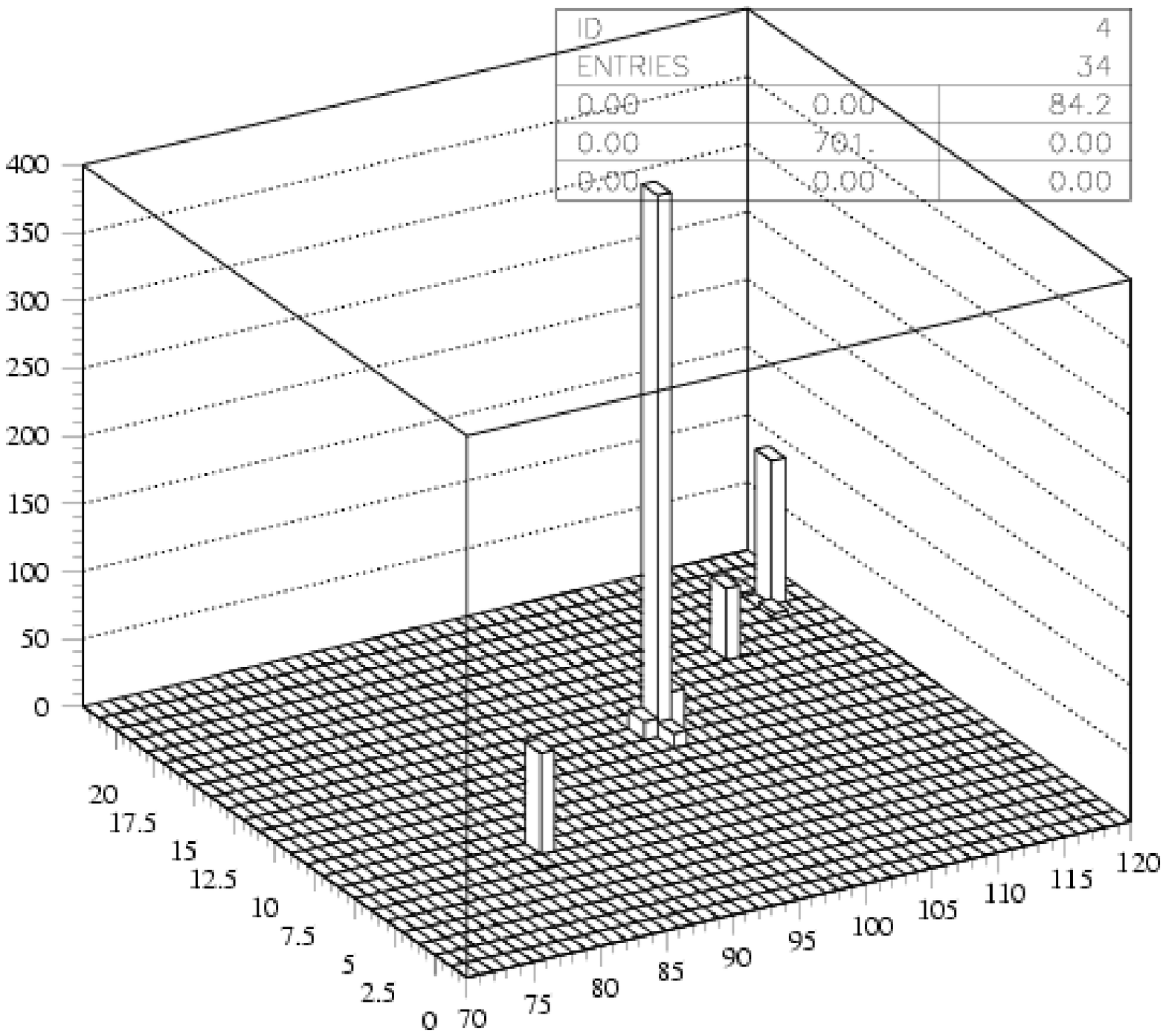}
\includegraphics*[width=0.48\textwidth,bbllx=40pt,bblly=20pt,bburx=540pt,bbury=470pt,angle=0,clip]{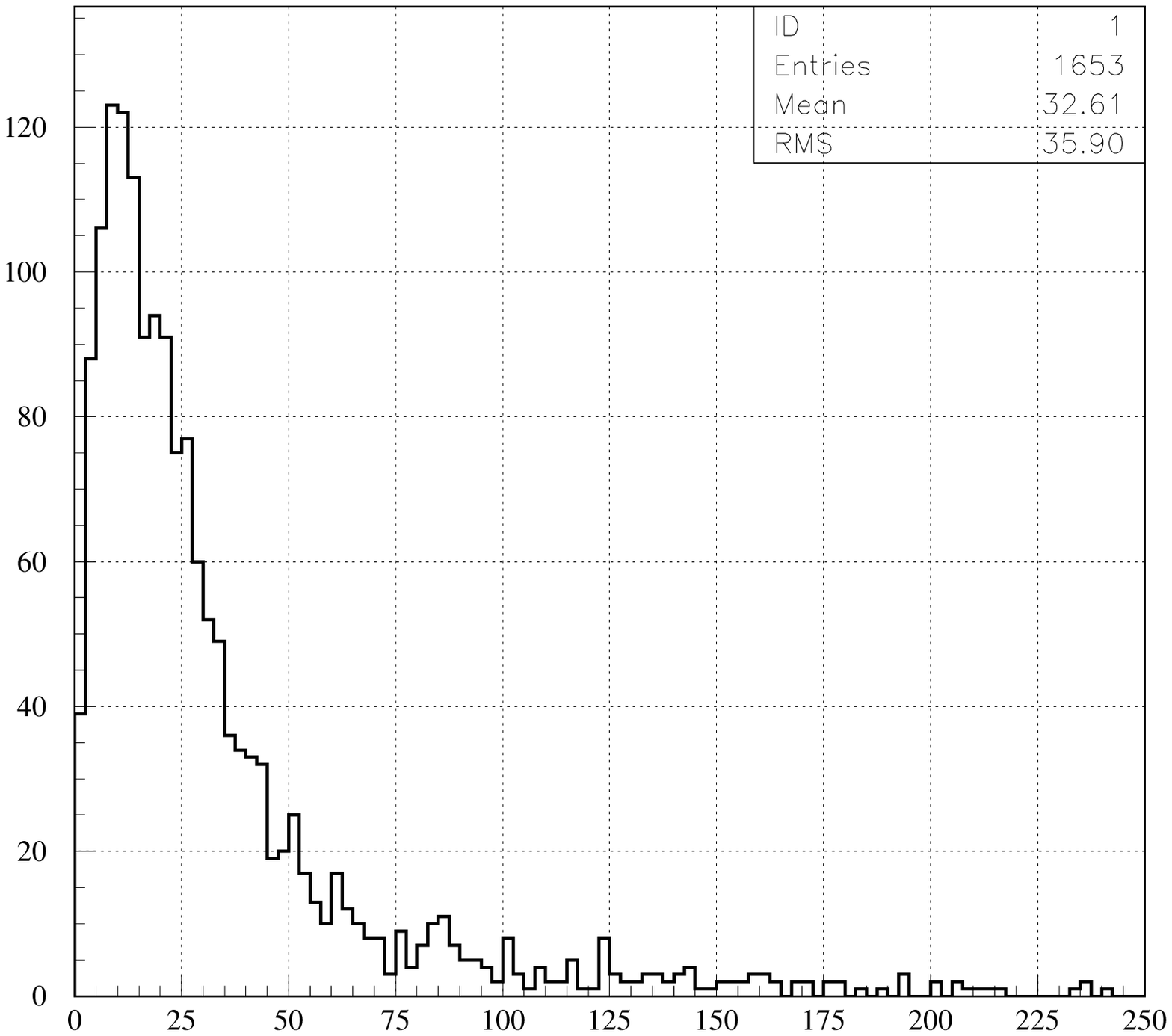}
\caption{ Left: Lego plot of the energy deposited 
by 34 $\gamma$'s in the alignment 
(energy in TeV on vertical axis, x, y in mm). 
Right:
 An example of the histogram of invariant mass for cluster A, 
with the mass of the $\pi^{0}$, the maximum gives 
an interaction distance near 100m
\label{fig1}}
\end{figure}
Inside the alignment, the four most energetic $\gamma$'s with respective energies, 300 TeV, 105 TeV, 75 TeV and 53 TeV are lying 
on a perfect straight line~\cite{jn01} with $r=0.9993$.
The invariant mass histograms for the different clusters
(Fig~\ref{fig1}) suggest an interaction level at about 100m above the Concorde, in contradiction with the simulation on the basis of standard physics requiring about 10km.
Such circumstance suggests a violent phenomena separating the valence quarks of the proton projectile, especially the valence diquark, which cannot be recombined with a quark of the sea, suppressing the leading baryon at the end of the collision.

One hypothesis is that the original rotating relativistic  string between the valence quark and the valence diquark becomes a more complex system with a secondary string (centered on the barycenter of the diquark) between the two quarks partners of the diquark. The maximal tension of the strings occurs when the quarks are at the largest distance from each other, i.e. when the 3 partons are on a common diameter which would be the axis of the fragmentation. The shorter mean free path of the diquark in the target nuclei could help the phenomena. As a consequence of the suppression of the leading particle, the maximum depth T$_{max}$
is expected to level off during one energy decade above the knee~\cite{jn03}
 and some typical behavior would have to be observed in EAS
( in muon electron dependance, in the age parameter versus size, different absorption length,enhanced steepeness of the most energetic $\gamma$'s  and hadrons spectra).
From other part, there have been few simulations with primary nuclei, and the recent proposition of
 a predominant $\alpha$ component suggests to examine more carefully the asymmetries in light nuclei collision, i.e. effects like the giant
 dipole resonance.
\section{Lateral distributions in giant EAS}
Some
functions are used in large surface arrays without reference to the
total size or to the age parameter, giving just an interpolation
between the detectors to evaluate the densities required for the
estimators at 600~m or 1000~m from axis.
The couple ($N_{e},s$) is especially
useful in the case of hybrid events~\cite{jf05}
 at the level of the registration;
 it can be derived from the fluorescence measurements
to start a minimisation on the densities recorded with the surface array and give a better
determination of the axis position with the hypergeometric functions\cite{jf05}:
\begin{eqnarray}
  f(x) = g(s)\ x^{s-a}(x+1)^{s-b}(1+dx)^{-c} \label{eq:JNC}
\end{eqnarray}
 with  $a=1.92$, $b=3.8$, $c=7.71$, $ d=0.00342$
(for distances r and densities $\Delta$,~ $r=xr_{0}$ and
$\Delta={ N \over r_{0}^{2} } f(r)$,$ r_{0}=36.8 m$, N being the total size.
 The calculation of the HG serial is replaced  by
 the approximation valid up to $\theta$=40$^{o}$ ~: $g(s)~=~-~0.19~+~0.969~s~-~0.468~s^{2}$.
  This relation works also for vme's
 with $a=1.94$,$b=3.92$, $c=2.87$, $d=0.00562$, $r_{0}=39.2$m, g(s) 
 being replaced by $g_{vme}(s)$~=~1~-~0.789 $s$~+~0.133$s^{2}$ and finally
$\Delta_{vme}$= $\phi_{1}(N_{e}/r_{0}^{2}) f(x)$ with $\phi_{1}=
0.47$ (agreement with experimental data in ref $[6]$).
\section{Primary spectra from classical and giant arrays }
The good extrapolation of the spectrum obtained in Akeno with the spectrum from
HIRES  Stereo is shown in Figure 2.
 In Akeno~\cite{nag92}, the densities were determined with a modest detector spacing ($
30$m or $100$m) and a specific lateral distribution, containing the age parameter
 was employed to localize the core and obtain the size N .
 The size N is converted directly to the primary energy with 
a relation in agreement with CORSIKA 
within $2\%$. The 20 km$^{2}$ array (Array 20) with 19 detectors,   
separated by about 1 km from each other, uses the distribution:
$\rho (r) = N \ C_{e} \ x^{-\alpha} (1+x)^{-(\eta-\alpha)} (1+{r\over{2000}})^{-
0.5}$
($C_{e}$  normalisation constant).
 This analytic description with a fixed value $\alpha$ = $1.2$, 
without reference to the age parameter is used to determine 
the axis position and to interpolate
 the value of the density at $600$m.

In contrast to the size conversion in Akeno, the scintillator response in terms of density S$_{600}$ is here converted to the primary energy following:
$\bar{E_{20}}(eV) =  2.0 10^{17} \times{( {S_{600} })}^{1.0}$
(S(r) is related to the electron and muon densities)

\begin{figure}[!h]
\begin{center}
\includegraphics[width=18pc,height=16pc]{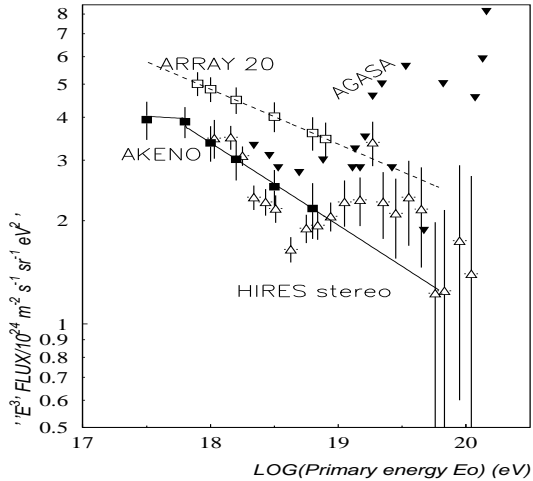}
\end{center}
\caption[Differential]{Differential primary spectrum for Array 1(full square), 
 Array 20(open square), AGASA(full triangle) 
and HIRES Stereo experiments(open triangle). 
Fits to Akeno(full line) and Array 20(dashed line) 
are from~\cite{nag92}: for the clarity of the graph, 
the error bars are not plotted for AGASA data.
\label{crsp1} }
\end{figure}

In place of the size spectrum, 
the $S_{600}$ differential spectrum in Array 20 is obtained 
taking an attenuation length
 $\Lambda_{600}$ in parallel to $\Lambda_{e}$ in Array 1 following:
\begin{equation}
  S_{600}(\Theta) =  
S_{600}(0)\times{exp(-{(t-t_{0})\over{\Lambda_{600}}}}) 
\label{eq:4}
\end{equation}
A constant value (also for AGASA)  $\Lambda_{600}$= $500g-cm^{-2}$ 
was  employed.
The most recent  values reported by AGASA~\cite{oli} are more close 
from the values of Akeno than the values of Array 20 (figure~\ref{crsp1}) ; 
the intensities of AGASA remain however larger than for Array 1 
in the overlapping energy region and exhibit a general excess 
by $30\%$ when compared to Hires Stereo
 data~\cite{oli}.
From our simulation data, we have derived the values 
of the attenuation length $\Lambda_{600}$ for different zenith angles:
 for small inclinations $\Theta \leq{30^{\circ}}$ the values 
of the attenuation length concerning
 proton primaries are quite more important
than the average value $\Lambda_{600}$= $500g-cm^{-2}$ used in AGASA. 
When the primary energy is increasing, the depth of the maximum 
becomes more and more close of the arrays in altitude, such 
as AUGER or AGASA :
 the conversion of inclined densities to $S_{600} (0)$ according 
to equation~\ref{eq:4} becomes poorly appropriate as the cascade is far 
from a stable absorption phase, especially for protons primaries.
 In the depth interval of about 5 radiation units following 
the maximum,  the absorption process is described by
the age parameter increasing in parallel from $1.0$ up to $1.2$,
%
 the lateral distribution around $600m$ from the axis becoming flatter.          
The increase of this flattening of the density distribution turns to a systematic overestimation (via  the vertical density from relation(4), the shower recorded may be classified in bins of larger energy).
Above $3.5~ 10^{19}$eV, a clear divergence in the discrepancies between AGASA and Hires Stereo appears rising from $150\%$ above $300\%$ at $6. 10^{19}$ eV.
 This may come again from the lateral distribution becoming flatter more rapidly  than the reduction of the total size : the net result is that the densities (at $600$ m) are $5-10\%$ larger in the bin $\Theta=20^{\circ}-30^{\circ}$than the vertical density
when the atmospheric depth separating the array and the shower maximum becomes lower than 3 cascade units. Some systematic errors could also enter in the axis localisation.

\section{Conclusions}
A large proportion of the alignments may be explained by fluctuations, however, the alignments observed in the stratosphere indicate the necessity of a more carefull analysis and the collection of new events, in the low stratosphere. $5000$ Hours could be available for a scientific payload during the certification of the Airbus A380. The flights carried at an altitude of $13.1$ km ($170g/cm^{-2}$) with 10 emulsion chambers, similar to those used in Concorde, would multiply by 100 the statistics of remarkable $\gamma$ ray families. This remains the most simple approach to the behavior of the valence quarks at energies close of the LHC energy range.
The present approach points out a better consistency
 between the spectra obtained by classical size measurements and Hires Stereo measurements, favourable to the GZK prediction.
The spectrum measured by the array KASCADE-Grande will be useful to improve the
 calibration of giant surface arrays.
\section*{Acknowledgments}
The author is indebted to the group of Lodz (J.Szabelski et al.) 
for usefull discussions and technical help to adapt the procedures
for producing the hard copy.

\end{document}